\documentclass[a4paper,11pt]{article}
\usepackage{pos}
\usepackage{graphicx}
\usepackage{xcolor}
\usepackage{subcaption} 

\title{Searching for very-high-energy electromagnetic counterparts to gravitational-wave events with the Cherenkov Telescope Array}
 \ShortTitle{Searching for VHE EM counterparts to GWs with CTA}

\author*[a,b]{Barbara Patricelli}
\author[c]{Alessandro Carosi}
\author[d]{Lara Nava}
\author[e]{Monica Seglar-Arroyo}
\author[f]{Fabian Sch\"ussler}
\author[b]{Antonio Stamerra}
\author[g]{Andrea Adelfio}
\author[f]{Halim Ashkar}
\author[h]{Andrea Bulgarelli}
\author[i]{Tristano Di Girolamo}
\author[h]{Ambra Di Piano}
\author[g]{Thomas Gasparetto}
\author[b]{Jarred Green}
\author[g]{Francesco Longo}
\author[j]{Ivan Agudo}
\author[k]{Alessio Berti}
\author[l]{Elisabetta Bissaldi}
\author[m]{Giancarlo Cella}
\author[i]{Antonio Circiello}
\author[d]{Stefano Covino}
\author[d]{Giancarlo Ghirlanda}
\author[n]{Brian Humensky}
\author[o]{Susumu Inoue}
\author[p]{Julien Lefaucheur}
\author[q]{Miroslav Filipovic}
\author[r,m]{Massimiliano Razzano}
\author[n]{Deivid Ribeiro}
\author[s]{Olga Sergijenko}
\author[h]{Giulia Stratta}
\author[p]{Susanna Vergani}
\forColl{CTA}

\affiliation[a]{European Gravitational Observatory, 
Via E Amaldi, I- 56021 Cascina (Pisa), Italy }

\affiliation[b]{INAF - Osservatorio Astronomico di Roma,  
Via Frascati 33, I-00078 Monte Porzio Catone (Rome), Italy }

\affiliation[c]{University of Geneva - D\'epartement de physique nucl\'eaire et corpusculaire, 
24 rue du G\'en\'eral-Dufour, 1211 Gen\`eve 4, Switzerland}

\affiliation[d]{INAF-Osservatorio Astronomico di Brera, 
Via E. Bianchi 46, I-23807 Merate (LC), Italy}

\affiliation[e]{LAPP, Univ. Grenoble Alpes, Univ. Savoie Mont Blanc, CNRS-IN2P3, 
9 Chemin deBellevue - BP 110, 74941 Annecy Cedex, France}

\affiliation[f]{IRFU, CEA, Universit\'e Paris-Saclay, 
Bât 141, 91191 Gif-sur-Yvette, France}

\affiliation[g]{INFN Sezione di Trieste and Universit\`a degli Studi di Trieste, 
Via Valerio 2 I-34127 Trieste, Italy}

\affiliation[h]{INAF - Osservatorio di Astrofisica e Scienza dello spazio di Bologna, 
Via Piero Gobetti9 3/3, I-40129 Bologna, Italy}

\affiliation[i]{INFN Sezione di Napoli and Universit\`a degli Studi di Napoli ``Federico II'',
 Complesso Universitario di Monte S. Angelo, 
Via Cintia, I-80126 Napoli, Italy}

\affiliation[j]{Instituto de Astrof\'isica de Andaluc\'ia-CSIC,
Glorieta de la Astronom\'ia s/n, 18008, Granada,Spain}

\affiliation[k]{INFN Sezione di Torino, 
Via P. Giuria 1, I-10125 Torino, Italy}

\affiliation[l]{INFN Sezione di Bari and Politecnico di Bari, 
via Orabona 4, I-70124 Bari, Italy}

\affiliation[m]{INFN - Sezione di Pisa, 
Largo B. Pontecorvo 3, I-56127 Pisa, Italy}

\affiliation[n]{Department of Physics, Columbia University, 
538 West 120th Street, New York, NY 10027, USA}

\affiliation[o]{RIKEN, Institute of Physical and Chemical Research, 
2-1 Hirosawa, Wako, Saitama,351-0198, Japan}

\affiliation[p]{LUTH, GEPI and LERMA, Observatoire de Paris, CNRS, PSL University, 
5 place Jules Janssen, 92190, Meudon, France}

\affiliation[q]{Western Sydney University, 
Locked Bag 1797, Penrith, NSW 2751, Australia}

\affiliation[r]{Dipartimento di Fisica, Universit\`a di Pisa, 
Largo B. Pontecorvo 3, I-56127 Pisa, Italy}

\affiliation[s]{Astronomical Observatory of Taras Shevchenko National University of Kyiv, 
3 Observatorna Street, Kyiv, 04053, Ukraine}

\emailAdd{barbara.patricelli@inaf.it}

\abstract{The detection of electromagnetic (EM) emission following the gravitational wave (GW) event GW170817 opened the era of multi-messenger astronomy with GWs and provided the first direct evidence that at least a fraction of binary neutron star (BNS) mergers are progenitors of short Gamma-Ray Bursts (GRBs). GRBs are also expected to emit very-high energy (VHE, > 100 GeV) photons, as proven by the recent MAGIC and H.E.S.S. observations. One of the challenges for future multi-messenger observations will be the detection of such VHE emission from GRBs in association with GWs. In the next years, the Cherenkov Telescope Array (CTA) will be a key instrument for the EM follow-up of GW events in the VHE range, owing to its unprecedented sensitivity, rapid response, and capability to monitor a large sky area via scan-mode operation. We present the CTA GW follow-up program, with a focus on the searches for short GRBs possibly associated with BNS mergers. We investigate the possible observational strategies and we outline the prospects for the detection of VHE EM counterparts to transient GW events.}

\FullConference{37$^{\rm{th}}$ International Cosmic Ray Conference (ICRC 2021)\\
		July 12th -- 23rd, 2021\\
		Online -- Berlin, Germany}

\begin{document}
\maketitle

\section{Introduction}
The joint observation of gravitational waves (GWs) from the binary neutron star (BNS) merger GW170817 \cite{gw170817} by Advanced LIGO \cite{LIGO} and Advanced Virgo \cite{Virgo} and of the short Gamma-Ray Burst GRB 170817A by \emph{Fermi}-GBM \cite{grb170817-fermi} and INTEGRAL \cite{grb170817-integral} marked the beginning of multi-messenger astronomy with GWs and provided the first direct evidence that at least a fraction of BNS mergers are progenitors of short GRBs \cite{grb170817}. The intense electromagnetic (EM) follow-up campaign performed after this joint detection also allowed to detect an optical/infrared counterpart to the GW event (the kilonova),  hosted in the galaxy NGC 4993; X-ray and radio counterparts have also been observed, and later identified as the GRB afterglow emission (see \cite{MMgw170817} and references therein). GRBs are also known to emit very-high-energy (VHE, E > 100 GeV) photons, as shown by the observations of GRB 190114C and GRB 160821B by MAGIC \cite{MAGIC-grb190114c,MAGIC-grb160821b} and GRB 180720B and GRB 190829A by  H.E.S.S. \cite{HESS-grb180720b,HESS-grb190829a}. A VHE EM follow-up of GW170817 was performed by H.E.S.S., that started the observations 5.3 hr after the GW event, but no EM counterpart was found \cite{HESS-grb170817a}. A search for a possible VHE EM counterparts has been performed also with HAWC, whose observations started $\sim$ 8 hr after GW170817, but no significant gamma-ray emission was found \cite{HAWC-grb170817a}.

In the coming years, the Cherenkov Telescope Array (CTA, \cite{CTA}) will play a fundamental role in the follow-up of GWs at VHE, thanks to its unprecedented sensitivity, its rapid slewing capabilities, and its large field-of-view (FOV). CTA will be composed of two arrays, one in the northern hemisphere and one in the southern hemisphere, which together will provide full-sky coverage; it will be an order of magnitude more sensitive and will have a greater energy coverage (from a few tens of GeV to above 100 TeV) with respect to current Imaging Atmospheric Cherenkov Telescopes (IACTs). The two arrays will consist of a combination of large (LST), medium (MST) and small (SST) size telescopes, covering different energy ranges: 20 GeV - 150 GeV, 150 GeV - 5 TeV and 5 TeV - 300 TeV, respectively. In the scheme of the CTA Key Science Project on transients, GW transient events are ranked as the highest priority ones to be studied \cite{CTAscience}. As a result, the GW    follow-up strategies with CTA, also investigated in previous studies (see, e.g. \citealp{patricelli2018,bartos2014,bartos2018,bartos2019,icrc2019}), are currently being defined.

In this paper, we present the EM follow-up program proposed for CTA and investigate the capability of CTA to detect VHE EM counterparts to GW transient events, based on detailed simulations of BNS mergers accompanied by short GRBs.

\section{The population of astrophysical sources}
\label{sec:population}
To investigate the capability of CTA to follow-up GW transient events and detect possible VHE EM counterparts, we simulate a catalog of short GRBs associated with GW signals from BNS mergers. This catalog of simulated BNS mergers and their GW detection was produced in expectation of the fourth observing run of current GW detectors (O4). Available in the public database \mbox{GWCOSMoS} \cite{GWcosmos}, the catalog is based on the work by \cite{patricelli2016,patricelli2018}. 
It has been built starting from a simulated, realistic ensemble of BNS merging systems evenly distributed in space up to a maximum distance of 500 Mpc, and contains only the events expected to be detected by Advanced LIGO and Advanced Virgo in O4; for these  events, the 2-dimensional GW skymaps are also available. 
We associate VHE emission to each simulated BNS merger, adopting the following empirical approach. According to the few available detections of VHE radiation from long GRBs \cite{MAGIC-grb190114c,HESS-grb180720b,HESS-grb190829a}, the VHE lightcurve looks similar to the soft X-ray lightcurve in terms of luminosity and temporal behaviour. Time-resolved spectral analysis of the brightest VHE event, GRB\,190114C, showed that the spectra are consistent with a power law (PL) with  photon index $\alpha\sim-2.2$ with no strong evidence for temporal evolution \cite{MAGIC-grb190114c}. 
Assuming that short GRBs also produce VHE radiation similarly to long GRBs (see \cite{MAGIC-grb160821b}), we simulate its   temporal and spectral properties as follows. Each BNS merger is assumed to successfully launch a relativistic jet, whose isotropic equivalent prompt emission energy $E_{\rm iso}$ follows the $E_{\rm iso}$ distribution of short GRBs derived in \cite{ghirlanda16}. The afterglow X-ray luminosity at 11\,h is assigned by adopting the $E_{\rm iso}-L_{\rm X,11h}$ correlation found for short GRBs in \cite{berger14}. We then assume $L_{\rm VHE,11h}\sim L_{\rm X,11h}$, allowing for a dispersion of 0.3\,dex. The full VHE lightcurve is built by assuming that it decays as a PL with an index extracted from the distribution of decay indices of X-ray afterglows of short GRBs; considering a sample of 22 short GRBs, we find that this distribution is well described by a Gaussian function with mean value $\langle\alpha_2\rangle=-1.45$ and $\sigma_{\alpha2}=0.48$. The initial Lorentz factor, assigned from a lognormal distribution centered around $\sim200$, determines the lightcurve peak time. Before the peak time, the light curve rises as $L_{\rm VHE}\propto t^2$. The off-axis lightcurve at viewing angle $\theta_{\rm view}$ is calculated following \cite{salmonson03} and \cite{lamb17}, and adopting a structured (Gaussian) jet with opening angle of the core taken from the distribution inferred in \cite{fong15}.

\section{CTA observations: exposure time versus latency time}
The VHE EM emission is expected to start shortly after the BNS merger, 
but the starting time of the EM follow-up observations typically doesn't coincide with the onset of the GRB emission. This is due to several factors: i) the latency needed to send the GW alert to astronomers (during the third observing run of Advanced LIGO and Advanced Virgo the typical latency was of the order of minutes, see \url{https://gracedb.ligo.org/superevents/public/O3/}, but in the future the latency could be shorter, see \url{https://emfollow.docs.ligo.org/userguide/early_warning.html}); ii) the time needed to point the telescopes in the region of the sky of interest (for instance, the slewing time of the LSTs is 30 s); iii) the uncertainty in the sky location of the GW event, that is typically of the order of tens/hundreds of square degrees (see, e.g., \cite{prospectsLVK}). This last point means that several consecutive pointings are needed to cover the GW localization region, and subsequently to pinpoint the sky location of the GRB (see, e.g., \cite{patricelli2018}). As a consequence, the exposure time needed to eventually detect the source can also vary, depending on the GRB luminosity and on the shape of its light curve.  

As a first step for our investigations, we estimate the exposure time needed to detect the GRBs in our simulated catalog with CTA as a function of the latency $t_0$ from the onset of the GRB emission and the starting of the observation of the sky region containing the source. Specifically, following the approach used in \cite{patricelli2018}, we estimate the exposure time needed to detect the source ($T_{\rm exp}$) as the time required to make a 5 $\sigma$ detection, i.e.  the time such that the following condition is fulfilled: 

\begin{equation}\label{eq1}
\int_{t_0}^{t_0+T_{\rm exp}} F(t) dt \geq F^s_{5\sigma}(T_{\rm exp}),
\end{equation}

where $F(t)$ is the GRB flux and $F^s_{5\sigma}(T_{\rm exp})$ is the minimum  fluence detectable by CTA for the exposure time $T_{\rm exp}$ at a $5\sigma$ significance level. This last term is computed for different exposure times using the ctools\footnote{http://cta.irap.omp.eu/ctools/; in this work we used the version 1.6.3.} function \emph{cssens}, with the Instrument Response Functions (IRFs) computed by the CTA consortium from detailed Monte Carlo simulations, in the so-called ``Production 3'' \cite{CTA-MM-2019}. The IRFs for the two arrays\footnote{In this work we considered the baseline array layouts:  \url{https://www.cta-observatory.org/science/cta-performance/}.} “North$\_$0.5h” and “South$\_$0.5h” have been generated by assuming a 30-minute observation of a point source situated at zenith angle 20$^\circ$. We assumed an offset between the position of the source and the center of the FOV of 1$^\circ$.

\begin{figure}[h!]
    \centering
    \begin{subfigure}{0.5\linewidth}
        \centering
        \includegraphics[width=0.9\textwidth]{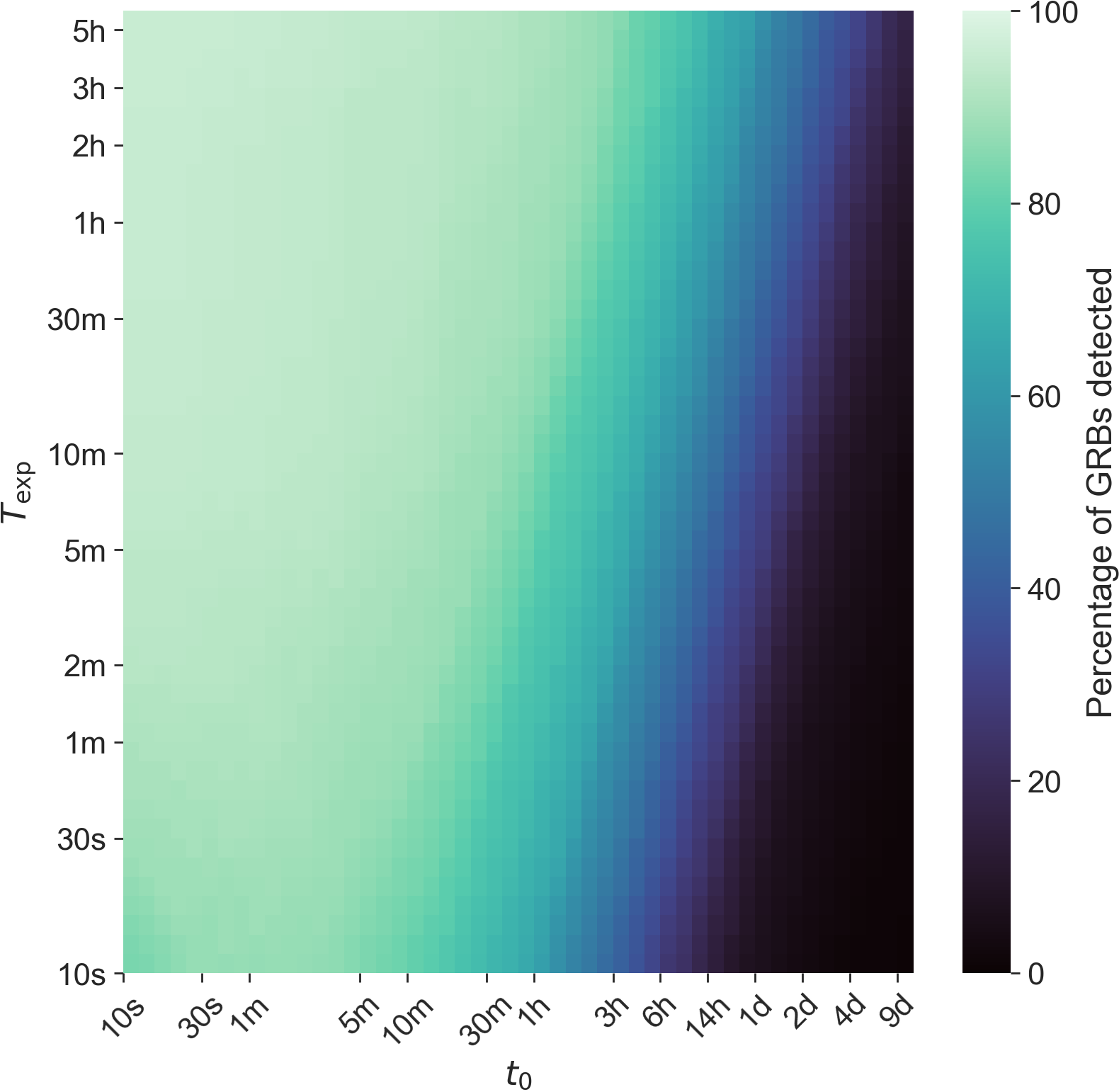}
       \caption{CTA South, z20$^{\circ}$, ($\theta_{\rm view} < 10^{\circ}$)}
    \end{subfigure}%
    \begin{subfigure}{0.5\linewidth}
        \centering
        \includegraphics[width=0.9\textwidth]{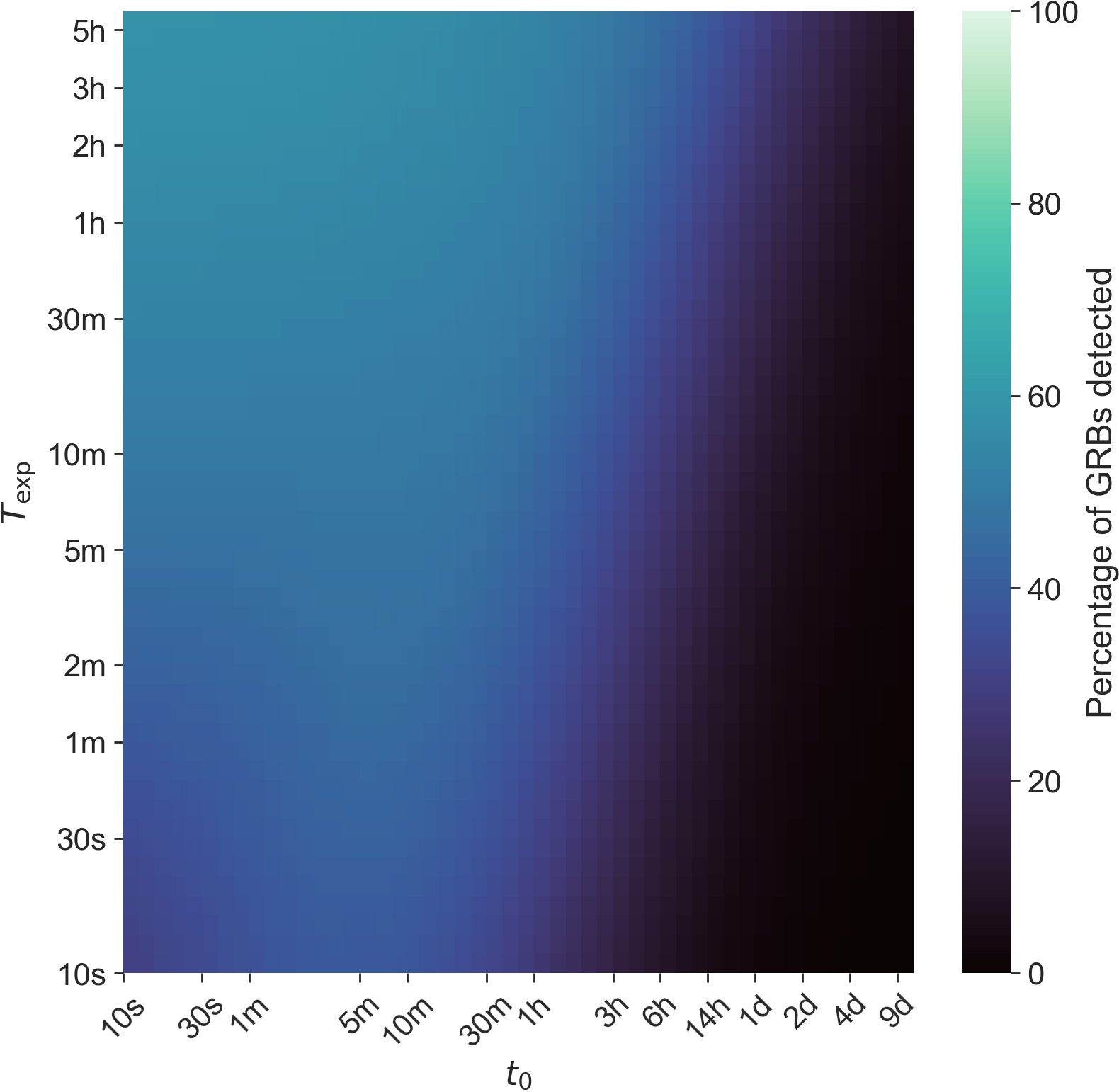}
       \caption{CTA South, z20$^{\circ}$, ($\theta_{\rm view} < 45^{\circ}$)}
    \end{subfigure}

    \begin{subfigure}{0.5\linewidth}
        \centering
        \includegraphics[width=0.9\textwidth]{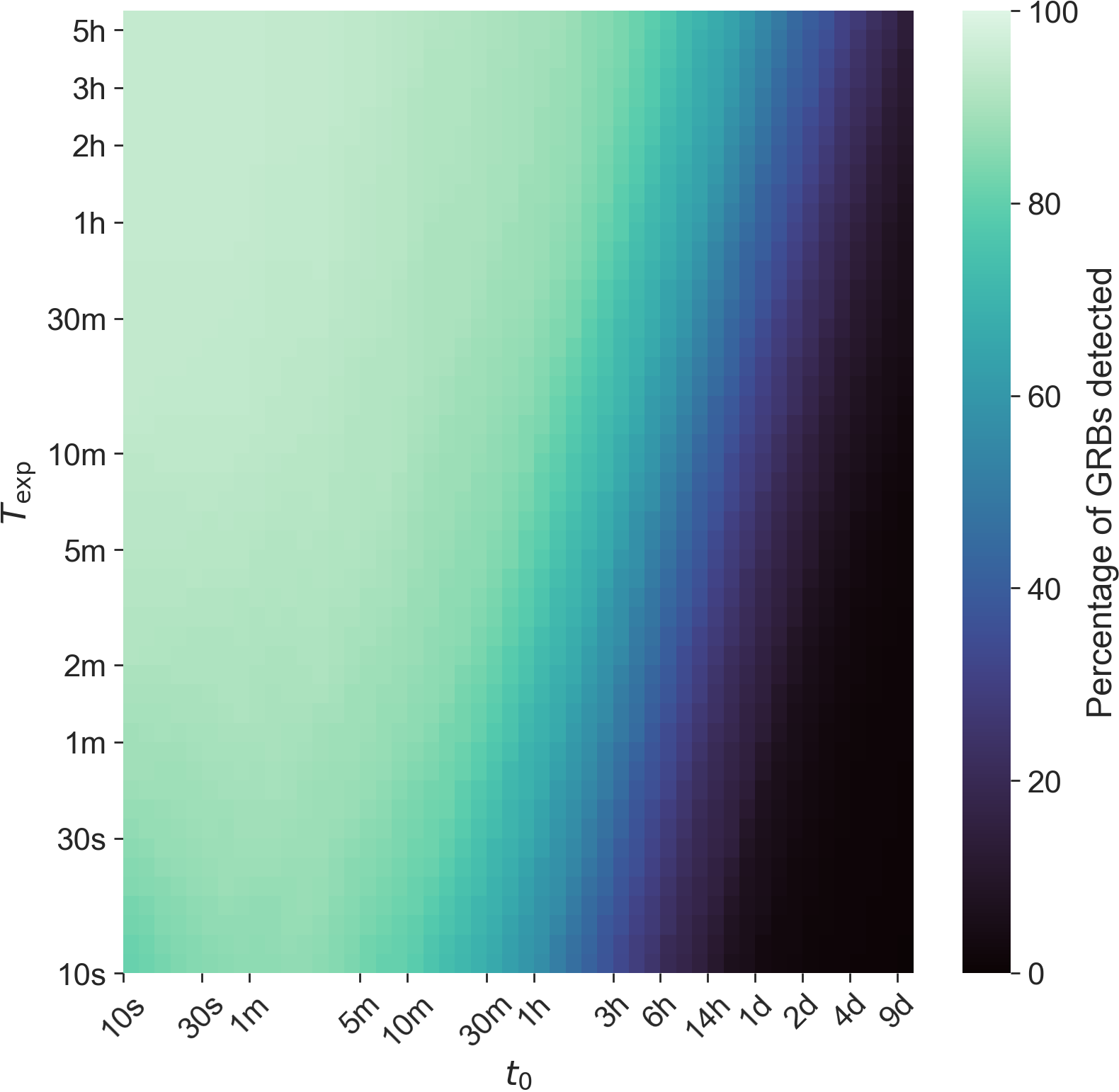}
       \caption{CTA North, z20$^{\circ}$, ($\theta_{\rm view} < 10^{\circ}$)}
    \end{subfigure}%
    \begin{subfigure}{0.5\linewidth}
        \centering
        \includegraphics[width=0.9\textwidth]{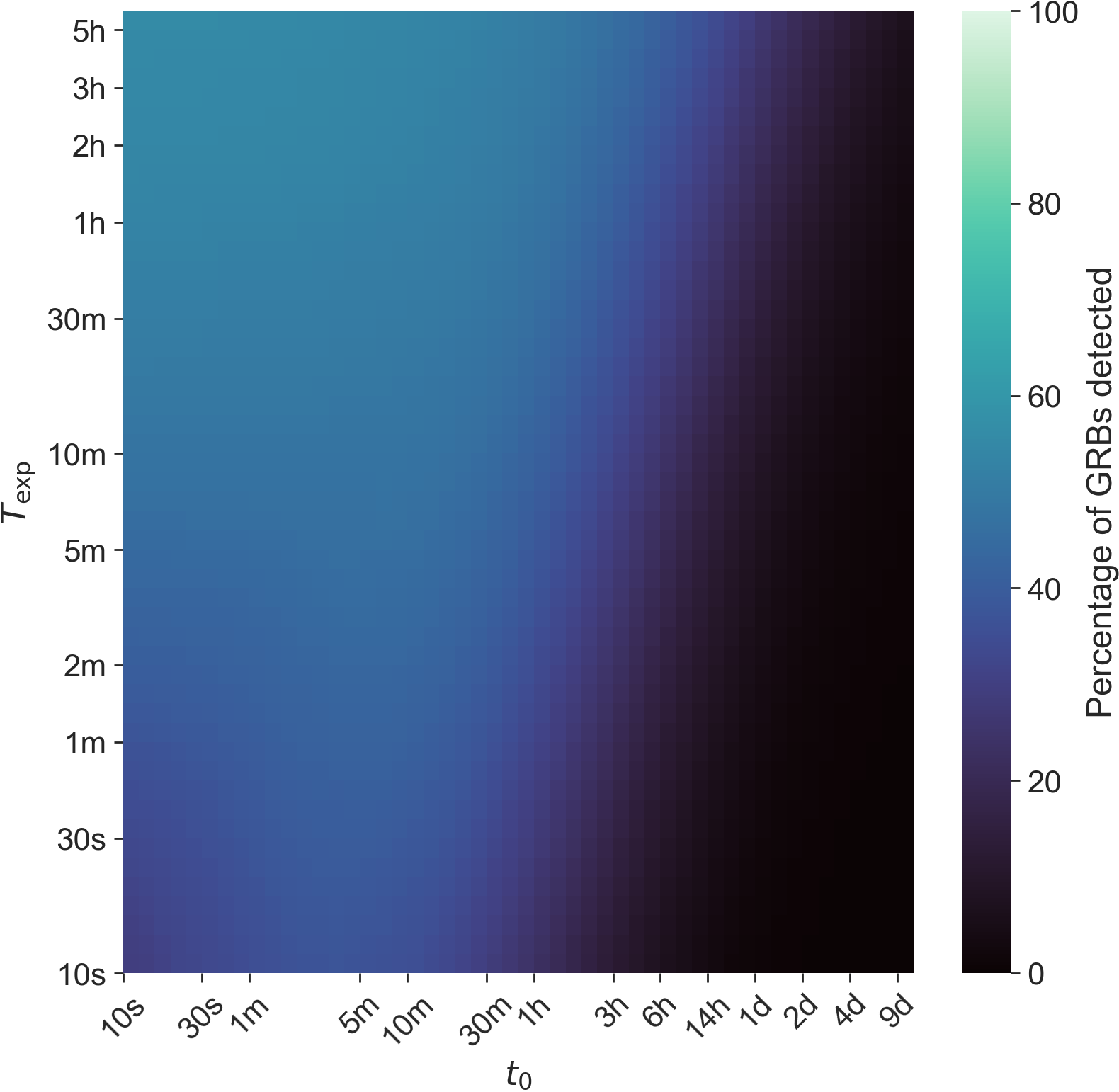}
       \caption{CTA North, z20$^{\circ}$, ($\theta_{\rm view} < 45^{\circ}$)}
    \end{subfigure}
    \caption{Percentage of short GRBs with $\theta_{\rm view} < 10^\circ$ (left panels) and $\theta_{\rm view} < 45^\circ$ (right panels) detected with CTA South (upper panels) and North (lower panels) with a given exposure time as a function of the delay time from the onset of the GRB emission and the starting of the observation of the sky region containing the source. A zenith angle of 20$^\circ$ has been assumed for all the GRBs.}
    \label{fig:percentage-GRBs}
\end{figure}

Figure \ref{fig:percentage-GRBs} shows the percentage of GRBs that could be detected by CTA for different exposure times, as a function of $t_0$, for GRBs seen at a viewing angle $\theta_{\rm view} < 10^\circ$ (these are mostly on-axis GRBs) and GRBs seen at $\theta_{\rm view} < 45^\circ$. When considering only on-axis GRBs and $t_0 \sim 10$~min, \mbox{$\sim$ 92 \%} of the sources can be detected with $T_{\rm exp}$ of the order of a few hours, by either CTA North of South alone; when considering a shorter delay of $t_0 \sim 30$~s, $\sim$ 94 \% of the sources can be detected with $T_{\rm exp} \leq$~30 minutes. When including off-axis GRBs for which  $\theta_{\rm view} < 45^\circ$, $\sim$ 54 \% of the sources can be detected within a few hours, considering $t_0  \sim$ 10~min; with the shorter delay $t_0 \sim 30$~s, $\sim$ 52 \% of these sources can be detected with $T_{\rm exp} \leq 30$~minutes.

\section{The CTA observational strategy}
The larger uncertainties on the source localization in GW events add an extra layer of complexity to the EM follow-up, i.e. to have a detection, the GW skymap region needs to be covered first. With the goal of studying the prospects for CTA and understanding how to maximise the chances of detecting a source, we study the most optimistic scenario: that is, to know \textit{a priori} the spectral and temporal evolution of the GRB, so the derived observation scheduling is optimal for each source (see, e.g., \cite{patricelli2018}). Then, by scanning the parameter space, we can derive the strategy which presents the best compromise for the input population described in Section \ref{sec:population}. 

\subsection{EM follow-up observations: the scheduler}
\label{scheduler}

Several observation scheduling algorithms have been developed to derive optimal pointing patterns which cover the largest total GW uncertainty region possible, an approach based on \cite{HESSrapid}. These algorithms are part of realistic observation scheduling simulations, which include the consideration of visibility conditions of both the North and South sites, i.e. darkness and moonlight conditions for each GW alert  
time. Other optimizations are performed regarding the prioritization of observations in low zenith angle conditions in order to achieve lower energy thresholds during observations.

Whereas in \cite{icrc2019} the main characteristics of this scheduler were introduced, we went a step further with these realistic simulations by considering the connection between the zenith angle evolution of the source, the computation of the exposure time from Eq.~\ref{eq1}, and the probability coverage maximisation in each iteration of the scheduler. This means that we are maximising our chances to detect the source, since the selected region is defined as being the one which encloses the highest GW source sky-position probability 
in each iteration of the observation strategy. The exposure is selected following Eq.~\ref{eq1} while considering zenith angle evolution as well, which is key for long exposure times.

\section{A test case}
The complete algorithm described in Subsection \ref{scheduler} is currently being used to investigate the GRB catalog associated with the simulated GW events from the GWCOSMoS database. Here we show one example corresponding to a simulated BNS merger located at a distance of $\sim$ 270 Mpc and whose GW sky localization area\footnote{Here we refer to the 90\% credible region, i.e., the area enclosing 90\% of the total posterior GW probability.} is $\sim$ 40 deg$^2$; the associated GRB is on-axis. 

\begin{figure}[h!]
    \centering
    \includegraphics[width=\textwidth]{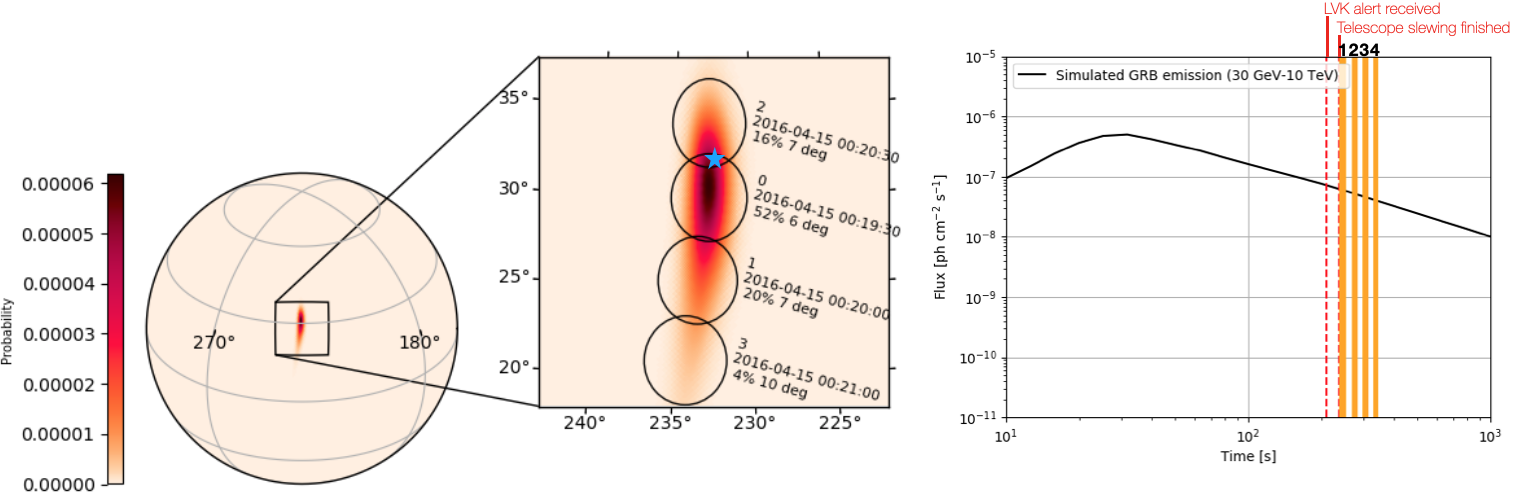}
    \caption{Observational scheduling of the test case from the GWCOSMoS catalog, for which four observations are scheduled and total of 92 \% of the uncertainty region is covered. The blue star marks the sky location of the BNS merger. For each observation, we quote the number, the time, the \% uncertainty region covered and the zenith angle.  The exposure times are selected such that a 5 $\sigma$ detection is achieved, which in this case correspond to $T_{\rm exp}$ = 10 s for all four observations. The injection time is 2016-04-15 00:15:30 UTC and CTA-North is selected due to the 90\% C.R. region. A total of 210 seconds are considered as latency for the GW alert to be received, 30 seconds for the first slewing of the telescopes, and 20 seconds for the final slewing before starting data acquisition. The simulated observation schedule is overlaid on the lightcurve of the test case GRB. We assumed a conservative FOV of 2.5$^\circ$ (see \cite{icrc2019}).}
    \label{fig:testcase}
\end{figure}

The injection time is 2016-04-15 00:15:30 UTC and has been selected so that the observations can be scheduled by at least one site, in this case CTA North. Four observations are scheduled, covering a 92\% of the uncertainty region in the localization of the source (see Fig. \ref{fig:testcase}). For this simulation, we consider the latency of the arrival of the GW alert to be 210 seconds, and the slewing time of the telescopes to be 30 seconds. The inter-slewing time to reach each position and the start of data acquisition between observations are estimated to be of 20 seconds. Thanks to this observation strategy, the source is covered and detected twice, in the first and third observation.

\section{Conclusions}
We have presented a study on the capability of CTA to detect VHE EM counterparts to GWs and discussed the possible observational strategies to follow-up GW transient events. We have shown that CTA represents a promising instrument to identify the VHE emission from GRBs associated with BNS mergers. Detailed estimates of the joint GW and VHE EM detection rates will be presented in a future work.

\acknowledgments
We gratefully acknowledge financial support from the agencies and organizations listed here: \href{http://www.cta-observatory.org/consortium_acknowledgments}{http://www.cta-observatory.org/consortium\_acknowledgments}.

\clearpage
\section*{Full Authors List: \Coll\ Consortium}

\parindent 0pt 

\scriptsize
\noindent

H.~Abdalla\textsuperscript{1}, H.~Abe\textsuperscript{2},
S.~Abe\textsuperscript{2}, A.~Abusleme\textsuperscript{3},
F.~Acero\textsuperscript{4}, A.~Acharyya\textsuperscript{5}, V.~Acín
Portella\textsuperscript{6}, K.~Ackley\textsuperscript{7},
R.~Adam\textsuperscript{8}, C.~Adams\textsuperscript{9}
% ...
\\

\noindent
1 : Centre for Space Research, North-West University, Potchefstroom, 2520, South Africa

2 : Institute for Cosmic Ray Research, University of Tokyo, 5-1-5, Kashiwa-no-ha, Kashiwa, Chiba 277-8582, Japan

3 : Pontificia Universidad Católica de Chile, Av. Libertador Bernardo O'Higgins 340, Santiago, Chile

4 : AIM, CEA, CNRS, Université Paris-Saclay, Université Paris Diderot, Sorbonne Paris Cité, CEA Paris-Saclay, IRFU/DAp, Bat 709, Orme des Merisiers, 91191 Gif-sur-Yvette, France

5 : Centre for Advanced Instrumentation, Dept. of Physics, Durham University, South Road, Durham DH1 3LE, United Kingdom

6 : Port d'Informació Científica, Edifici D, Carrer de l'Albareda, 08193 Bellaterrra (Cerdanyola del Vallès), Spain

7 : School of Physics and Astronomy, Monash University, Melbourne, Victoria 3800, Australia

8 : Laboratoire Leprince-Ringuet, École Polytechnique (UMR 7638, CNRS/IN2P3, Institut Polytechnique de Paris), 91128 Palaiseau, France

9 : Department of Physics, Columbia University, 538 West 120th Street, New York, NY 10027, USA

% ...


%
%

\end{document}